\documentclass[showpacs,amsmath,twocolumn,showkeys,prl]{revtex4-1}
\usepackage{dcolumn}    
\usepackage{bm}         
\usepackage{ifpdf}
\usepackage{amssymb,lineno,amsfonts}
\usepackage{graphicx}   
\usepackage{bbm}
\usepackage{mathrsfs}
\usepackage{upgreek}
\usepackage{epstopdf}
\usepackage{setspace}
\usepackage{hyperref}
\usepackage[matrix,frame,arrow]{xypic}
\usepackage{float}
\usepackage{natbib}
\usepackage{color} 
\usepackage{subfigure}
\usepackage{lipsum}
\definecolor{med-blue}{RGB}{25,25,112} 
\hypersetup{colorlinks, linkcolor={blue},citecolor={blue}, urlcolor={blue}}
\newcommand{\ket}[1]{\vert{#1}\rangle}

\newcommand{\inpr}[2]{\langle{#1}\vert{#2}\rangle}

\newcommand{\tr}{\mathrm{Tr}}

\begin{document}
	\title{Push-Pull Optimization of Quantum Controls}
	\author{Priya Batra}
	\email{priya.batra@students.iiserpune.ac.in}
	\author{V. R. Krithika}
	\email{krithika$_$vr@students.iiserpune.ac.in}
	\author{T. S. Mahesh}
	\email{mahesh.ts@iiserpune.ac.in}
	\affiliation{Department of Physics and NMR Research Center,\\
		Indian Institute of Science Education and Research, Pune 411008, India}

	\begin{abstract}
{Optimization of quantum controls to achieve a target process is centered around an objective function comparing the a realized process with the target. 
We propose an objective function that incorporates not only the target operator but also a set of its orthogonal operators, whose combined influences  lead to an efficient exploration of the parameter space,  faster convergence, and extraction of superior solutions.  The \textit{push-pull} optimization, as we call it, can be adopted in various quantum control scenarios.  We describe adopting it to a gradient based and a variational-principle based approaches.  Numerical analysis of quantum registers with up to seven qubits reveal significant benefits of the push-pull optimization.  Finally, we describe 
applying the push-pull optimization to prepare a long-lived singlet-order in a two-qubit system using NMR techniques.}
	\end{abstract}
	
	\keywords{Optimal control theory, gradient optimization, variational approach, nuclear magnetic resonance}
	
	\maketitle
\textit{Introduction:}
Optimal control theory finds applications in diverse fields such as finance, science, engineering, etc. \cite{bryson2018applied,pontryagin2018mathematical}.  
Quantum optimal control has also gained significant attention over the last several years \cite{werschnik2007quantum,dong2010quantum} and is routinely used in robust steering of quantum dynamics as in chemical kinetics \cite{brumer1989coherence,petzold1999model}, spectroscopy \cite{tannor1985control,zhu1998rapidly,nielsen2007optimal}, quantum computing \cite{palao2002quantum,doria2011optimal}, and many more.
Here we focus on optimization of quantum controls to either transfer from one state to another, henceforth called state control, or to realize a target unitary evolution, henceforth called gate control.  Relevant numerical techniques fall into several categories including: stochastic-search methods such as strongly modulating pulses \cite{fortunato2002design};  gradient based approaches such as gradient ascent pulse engineering (GRAPE) \cite{khaneja2005optimal,de2011second} and gradient optimization of analytical control (GOAT) \cite{machnes2018tunable}; variational-principle based Krotov optimization \cite{krotov2008quantum,maximov2008optimal,reich2012monotonically}; truncated basis approach such as chopped random basis  optimization (CRAB) \cite{caneva2011chopped,sorensen2018quantum}; genetic algorithm enabled bang-bang controls \cite{bhole2016steering,khurana2017bang}; and machine learning based approaches \cite{chen2013fidelity, zhang2019reinforcement}. These control schemes have been implemented on various quantum architectures such as NMR \cite{fortunato2002design,vandersypen2005nmr,nielsen2007optimal,bhole2016steering}, NV centers \cite{dolde2014high}, ion trap \cite{singer2010colloquium}, superconducting qubits \cite{shim2016semiconductor}, magnetic resonance imaging \cite{vinding2012fast}, cold atoms \cite{doria2011optimal} etc.

An objective function evaluating the overlap of the realized process with the target process is at the core of an 
optimization algorithm and therefore should be chosen carefully \cite{chakrabarti2007quantum,pechen2011there}. Here we propose a hybrid objective function that not only depends on the target operator, but also on a set of orthogonal operators.
One may think of control parameters being \textit{pulled} by the target operator as well as \textit{pushed} by the orthogonal operators.
Accordingly, we refer to this method as \textit{Push-Pull Optimization of Quantum Controls} (PPOQC).   We describe adopting PPOQC for GRAPE and Krotov algorithms and demonstrate its superior convergence over the standard \textit{pull-only} methods. 
We also experimentally demonstrate the efficacy of PPOQC in a NMR quantum testbed by preparing long-lived singlet-order.

\begin{figure}
	\centering
	\includegraphics[trim=0cm 3.8cm 0cm 3.6cm,width=6.5cm,clip=]{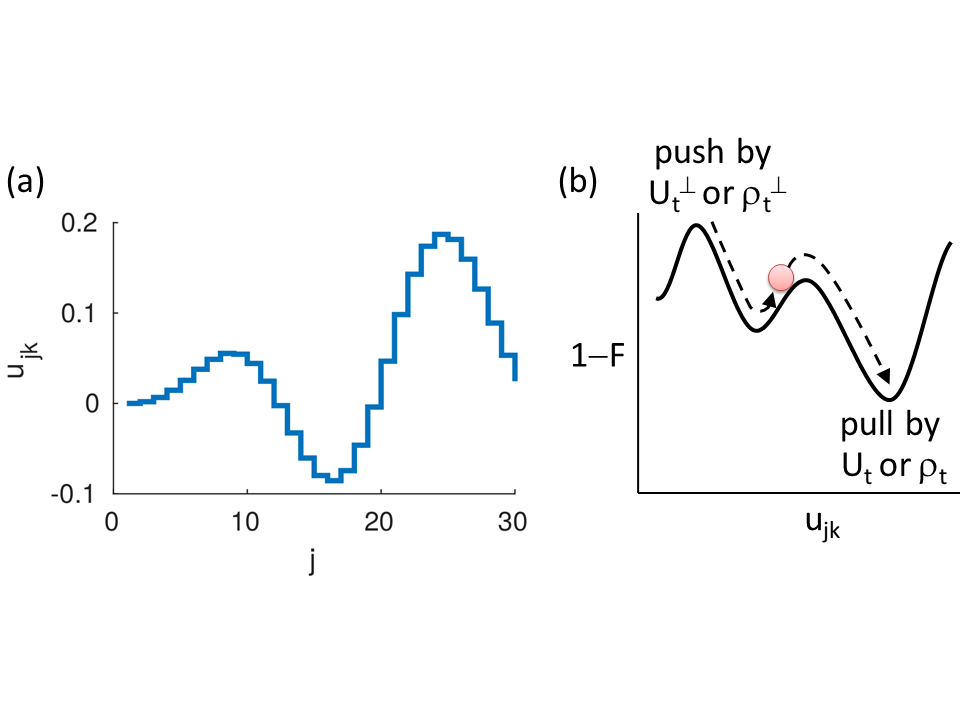}
	\caption{(a) Piecewise-constant control parameter $u_{jk}$ versus the segment number $j$.  (b) Infidelity $1-F$ versus $u_{jk}$.  }
	\label{scheme}
\end{figure}

\textit{The optimization problem:}
Consider a quantum system with an internal or fixed Hamiltonian $H_0$ and a set of $M$ control operators $\{A_k\}$ leading to the full time-dependent Hamiltonian
\begin{equation}
H(t) = H_{0} + \sum_{k=1}^{M} u_{k}(t) A_{k},
\end{equation}
where control amplitudes $u_k(t)$ are amenable to optimization.
The propagator for a control sequence of duration $T$ is  
$D\exp\left(-i\int_{0}^{T}H(t')dt'\right)$,
where $D$ is the Dyson time-ordering operator.  The standard approach to simplify the propagator is via piecewise-constant control amplitudes with $N$ segments each of duration $\tau$ (see Fig. \ref{scheme}(a)). In this case, the overall propagator is of the form
$U_{1:N} = U_N U_{N-1} \cdots U_2 U_1$, where
$U_{j} = \exp(-iH_j\tau)$ is the propagator for the $j$th segment
and 
$H_j = H_0 + \sum_{k=1}^{M} u_{jk} A_k$.
Our task is to optimize the control sequence $\{u_{jk}\}$ depending on the following two kinds of optimizations: \newline
\noindent (i) \textit{Gate control} (GC): Here the goal is to achieve an overall propagator (gate) $U_t$ that is independent of the initial state. This is realized by maximizing
the gate-fidelity
	\begin{equation}
	F(U_{1:N},U_t) =  \left|\inpr{U_t}{U_{1:N}}\right|^{2} = 
	 \left|\tr\{U_t^{\dagger}U_{1:N}\}\right|^{2}.
	\end{equation} 
\noindent (ii) \textit{State control} (SC): Here the goal is to drive a given initial state $\rho_0$ to a desired target state $\rho_t$.  This can be achieved by maximizing the state-fidelity
	\begin{equation}
F(\rho_{1:N},\rho_t) = \inpr{\rho_t}{\rho_{1:N}}= \tr\left\{\rho_t \rho_{1:N}\right\},
	\end{equation}
	where $\rho_{1:N} = U_{1:N}\rho_0U_{1:N}^\dagger$. 

In practice, hardware limitations impose bounds on the control parameters ${\{u_{jk}\}}$ and therefore it is desirable to minimize the overall control resource $r_k = \sum_{j} u_{jk}^2$.  To this end, we use the performance function 
$J = F - \sum_{k=1}^{M} \lambda_k r_k$,
where $\lambda_k$ are penalty constants.

\textit{PPOQC}:
Be it gate control or state control, for a $d$-dimensional target operator, we can efficiently setup $d-1$ orthogonal operators via Gram-Schmidt orthogonalization procedure \cite{shankar2012principles}.  The target operator \textit{pulls} the control-sequence towards itself, whereas the orthogonal operators \textit{push} it away from them (see Fig. \ref{scheme}(b)).
We define the \textit{push} fidelities as 
\begin{eqnarray}
\mbox{GC:}~~ F_o(U_{1:N},\{V_l\}) &=& \frac{1}{L} \sum_{l=1}^{L} F(U_{1:N},V_l) ~~\mbox{and} \nonumber \\ 
\mbox{SC:}~~  F_o(\rho_{1:N},\{R_l\}) &=& \frac{1}{L}\sum_{l=1}^{L} F(\rho_{1:N},R_l),
\end{eqnarray}
where $\{V_l\}$ and $\{R_l\}$ are $L \le d-1$ orthogonal operators such that $F(U_t,V_l) = 0$ and $F(\rho_t,R_l) = 0$.
Of course, $d$ increases exponentially with the system size, but as we shall see later, a small subset of $L$ orthogonal operators can bring about a substantial advantage.
Also, note that for a given target operator, the set of orthogonal operators is not unique and can be generated randomly and efficiently in every iteration.  
We define the \textit{push-pull} performance function 
\begin{equation}
J_{PP} = F - \alpha F_o  - \sum_{k=1}^{M} \lambda_k r_k,
\label{ppfid}
\end{equation}
where $-1 \le \alpha \le 1$ is the \textit{push} weight. 
In the following, we describe incorporating PPOQC into two popular optimal quantum control methods.

GRAPE \textit{optimization: }
Being a gradient based approach, it involves an efficient calculation of the maximum-ascent direction \cite{khaneja2005optimal}.  While it is sensitive to the initial guess and looks for a local optimum,  it is nevertheless simple, powerful, and popular.  The algorithm iteratively updates  control parameters $\{u_{jk}\}$ in the direction of gradient 
$g_{jk}^{(i)} = \partial J^{(i)}/\partial u_{jk}^{(i)}$:
\begin{eqnarray}
\mbox{GC:}~~ g_{jk}^{(i)}(U_t) &=& 2\tau ~\mathrm{Im}\{\inpr{P_{j}}{ A_{k}U_{1:j}}\inpr{U_{1:j}}{P_j} \} 
\nonumber \\
\mbox{SC:}~~ g_{jk}^{(i)}(\rho_t) &=& -i \tau \inpr{\tilde{\rho}_j}{[A_{k},\rho_{1:j}]},
\end{eqnarray}
where $i$ denotes iteration number,
$P_j = U_{j+1:N}^\dagger U_t$
and
$\tilde{\rho}_j = U_{j+1:N}^\dagger \rho_t U_{j+1:N}$
\cite{khaneja2005optimal}.
Collective updates $u_{jk}^{(i+1)} = u_{jk}^{(i)} + \epsilon g_{jk}^{(i)}$
after iteration $i$ on all the segments  with a suitable step size $\epsilon$, proceeds with monotonic convergence. 

\textit{Push-pull GRAPE (PP-GRAPE):}
Using Eq. \ref{ppfid} we recast the gradients as
\begin{eqnarray}
\mbox{GC:}~~  G_{jk}^{(i)}(U_t,\{V_l\}) &=& g_{jk}^{(i)}(U_t) - \frac{\alpha}{L}\sum_{l=1}^{L}g_{jk}^{(i)}(V_l)
\nonumber ~~\mbox{and} \\
\mbox{SC:}~~  G_{jk}^{(i)}(\rho_t,\{R_l\}) &=& g_{jk}^{(i)}(\rho_t) -
\frac{\alpha}{L}\sum_{l=1}^{L}g_{jk}^{(i)}(R_l),
\end{eqnarray}
and the update rule as
$u_{jk}^{(i+1)} = u_{jk}^{(i)} + \epsilon G_{jk}^{(i)}$.
The revised gradients form the basis of PP-GRAPE.

\begin{figure*}
	\centering
	\includegraphics[trim=3cm 1cm 2cm 0.8cm,width=17cm,clip=]{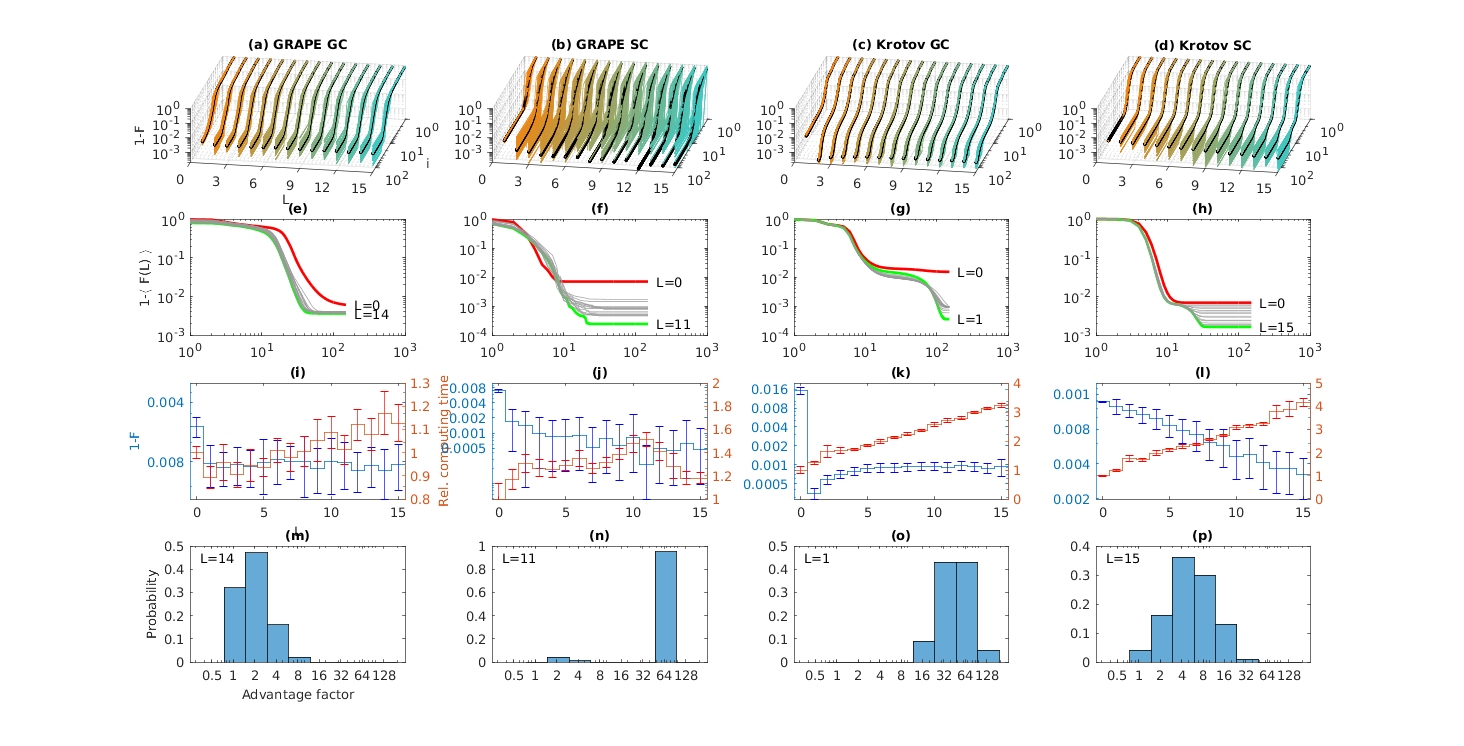}
	\caption{(a-d) Infidelity $1-F$ of two-qubit controls versus iteration number ($i$) and number ($L$) of orthogonal operators for GRAPE and Krotov as indicated.  Black lines represent mean infidelities. (e-h) Mean infidelity versus $i$. Curves for $L=0$ (red) and for $L$ leading to the maximum final fidelity (green) are highlighted. (i-l) Mean final infidelity
    (left axis) and relative computing time (right axis) versus $L$.  Error bars represent one standard deviation.  (m-p) Probability versus  advantage factor.  }
	\label{grkr}
\end{figure*} 

\textit{Krotov optimization:}
Based on  variational-principle, this method aims for the global optimum \cite{krotov1995global}. Here the performance function is maximized with the help of an appropriate Lagrange multiplier $B_j$.  One sets up a Lagrangian of the form \cite{nielsen2007optimal},
\begin{equation}
{\cal L} =  F - \sum_{k=1}^{M} \lambda_k r_k -\sum_{j=1}^{N} \mbox{Re}\left \langle
B_{j}\left \vert \frac{d}{dt}+iH_{j} \right \vert U_{0:j} \right \rangle,
\end{equation}
where the first two terms are same as the performance function $J$,
and looks for a stationary point satisfying
$\frac{\partial {\cal L}}{\partial F} = 0, ~~
\frac{\partial {\cal L}}{\partial u_{jk}} = 0, ~~ \mbox{and,} ~~
\frac{\partial {\cal L}}{\partial B_{j}} = 0$.
The second differential equation leads to
$u_{jk} = \frac{1}{\lambda_k}\mbox{Im}\inpr{B_j}{A_k U_{0:j}}$,
and the last differential equation constrains evolution according to the Schr\"{o}dinger equation
$\dot{B}(t) = -i H(t) B(t)$.

At every iteration $i$, the Krotov algorithm evaluates the control sequence
$\{u_{jk}^{(i)}\}$ as well as its co-sequence $\{\widetilde{u}_{jk}^{(i)}\}$.
Starting with a random guess $\{u_{jk}^{(0)}\} = \{\widetilde{u}_{jk}^{(0)}\}$, forward propagation of the sequence $\{u_{jk}^{(0)}\}$ gives $U_{1:j}$ and  backward propagation of the co-sequence $\{\widetilde{u}_{jk}^{(0)}\}$ from the boundary $B_N = \partial F /\partial U_{1:N}$ leads to $B_j$.
Specifically,  
\begin{eqnarray}
\mbox{GC:}~~ 
B_N  &=&  \inpr{U_t}{U_{0:N}} U_t 
\nonumber \\
\mbox{SC:}~~ 
B_N &=& \rho_t U_{0:N} \rho_0 + \kappa U_{0:N}.
\label{B}
\end{eqnarray}
Here $U_{0:N} = U_0U_{1:N}$, $U_0 = \mathbbm{1}$, and $\kappa$ is a positive constant that ensures the positivity of fidelity.
Back propagating the co-sequence, we obtain
\begin{equation}
B_{j} = \widetilde{U}_{j+1}^\dagger \cdots \widetilde{U}_{N-1}^\dagger \widetilde{U}_N^\dagger B_N,
\label{Bj}
\end{equation}
where
$\widetilde{U}_j = \exp(-i\widetilde{H}\tau)
~~ \mbox{and} ~~ \widetilde{H}_j = H_0 + \sum_{k=1}^{M} \widetilde{u}_{jk} A_k$.
Now the sequence $\{u_{jk}^{(i)}\}$ is updated according to
\begin{equation}
u_{jk}^{(i)} = (1-\delta)\widetilde{u}_{jk}^{(i-1)} + \frac{\delta}{\lambda_k}\mbox{Im}
\inpr{B_j^{(i-1)}}{A_k U_{0:j-1}^{(i)}}
\end{equation}
and propagator $U_{0:j}^{(i)}$ is evaluated. 
Iterating the last two steps delivers propagators $U_{0:1}^{(i)}, U_{0:2}^{(i)}, \cdots, U_{0:N}^{(i)}$. The terminal Lagrange multiplier $B_N^{(i)}$ is evaluated using the Eq. \ref{B}.  To setup the co-sequence $\{\widetilde{u}_{jk}^{(i)}\}$ we first evaluate the terminal control $\widetilde{u}_{Nk}$ using
\begin{equation}
\widetilde{u}_{jk}^{(i)} = (1-\eta)u_{jk}^{(i)} + \frac{\eta}{\lambda_k}\mbox{Im}
\inpr{B_j^{(i)}}{A_k U_{0:j}^{(i)}}
\label{tildeujk}
\end{equation}
with $j=N$. 
The Lagrange multiplier $B_{N-1}^{(i)} = \widetilde{U}_N^\dagger B_N^{(i)}$ is now evaluated by back-propagating with the updated amplitude $\widetilde{u}_{Nk}^{(i)}$.  
Iterating the last two steps updates the whole co-sequence $\{\widetilde{u}_{jk}^{(i)}\}$.  
The algorithm is continued until the desired fidelity is reached.

\textit{Push-pull Krotov (PP-Krotov):}
Here we use $L$ additional co-sequences  $\{\widetilde{v}_{jkl}^{(i)}\}$ corresponding to orthogonal operators $\{V_l\}$ or $\{R_l\}$.
Terminal Lagrange multipliers $\{C_{Nl}\}$ are obtained using similar equations as in Eq. \ref{B}:
\begin{eqnarray}
\mbox{GC:}~~ 
C_{Nl}  &=&  \inpr{V_l}{U_{0:N}} V_l 
\nonumber \\
\mbox{SC:}~~ 
C_{Nl} &=& R_l U_{0:N} \rho_0 + \kappa U_{0:N}.
\end{eqnarray}
The intermediate Lagrange multipliers $C_{jl}$ are evaluated by back-propagating $C_{Nl}$ in a similar way as described in Eq. \ref{Bj}, but by replacing the target operator with orthogonal operator $V_l$ (or $R_l$).
Revised update rule is
\begin{eqnarray}
u_{jk}^{(i)} &=& (1-\delta)\widetilde{u}_{jk}^{(i-1)} + \frac{\delta}{\lambda_k}\mbox{Im}
\inpr{B_j^{(i-1)}}{A_k U_{0:j-1}^{(i)}} \nonumber \\
&& + \frac{\alpha \delta}{L} \sum_{l=1}^L \left[ \widetilde{v}_{jkl} - \frac{1}{\lambda_k}
\inpr{C_{jl}^{(i-1)}}{A_k U_{0:j-1}^{(i)}}\right],
\end{eqnarray}
where
$\widetilde{v}_{jkl}^{(i)} =
\frac{\alpha\eta}{L} 
\left[
u_{jk}^{(i)} -
\frac{1}{\lambda_k}\sum_{l=1}^L 
\mbox{Im}
\inpr{C_{jl}^{(i)}}{A_k U_{0:j}^{(i)}}
\right]$ and $\alpha$ is the push weight as in Eq. \ref{ppfid}.  We now proceed to numerically analyze PPOQC performance.

\textit{Numerical analysis:}
Results of PPOQC analysis in a model two-qubit Ising-coupled system are summarized in Fig. \ref{grkr}.  For GC, we use CNOT gate as the target, while for SC, the task is a transfer from $\ket{00}$ state to singlet state $\ket{S_0} = (\ket{01}-\ket{10})/\sqrt{2}$.  In each case, we use a fixed set of one hundred random guess-sequences.  PP-GRAPE and PP-Krotov algorithms were run for various sizes of orthogonal sets ($L\in[1,15]$ with push weight $\alpha=0.2$) and compared with the pull-only ($L=0$) results (Fig. \ref{grkr}(a-d)).  
PPOQC outperformed the pull-only algorithms in terms of the mean final fidelity in all the cases (Fig. \ref{grkr} (e-h)).  
More importantly, while the pull-only fidelities tend to saturate by settling into local minima, the push-pull trials appeared to explore larger parameter space and thereby extracted solutions with better fidelities.
While the computational time for PP-GRAPE is weakly dependent on $L$, we find a slow but linear increase in the case of PP-Krotov (Fig. \ref{grkr} (i-l)).  
To quantify the advantage of PPOQC over the standard algorithms, we define the advantage factor $(1-F(L=0))/(1-F(L_\mathrm{best}))$, where $L_\mathrm{best}$ corresponds to the one with maximum mean of final-fidelity (Fig. \ref{grkr} (m-p)).  
In all the cases PPOQC ($L \ge 1$) resulted in  superior convergences than the standard \textit{pull-only} ($L=0$) algorithms.  In particular,  PP-Grape SC and PP-Krotov GC reached advantage factors up to 64, while PP-Krotov SC reached up to 16.  Only in PP-Grape GC, the advantage factor was modest 2.

\begin{figure}
	\centering
	\includegraphics[trim=3cm 9.3cm 3cm 9.8cm,width=8cm,clip=]{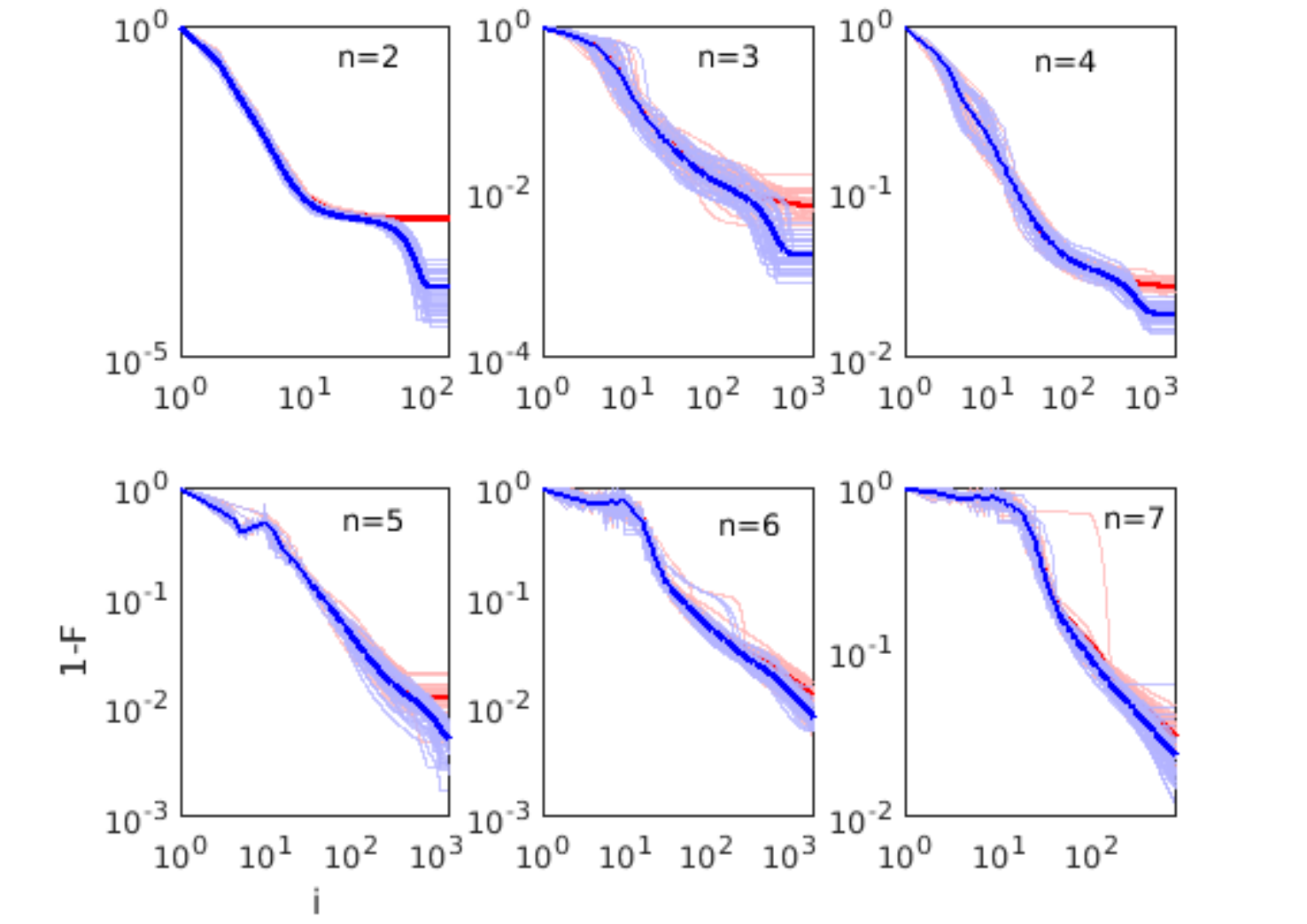}
	\caption{Infidelities for 40 random guesses (thin lines) and their mean (thick lines) versus iteration number $i$ with Krotov (red) and with PP-Krotov (blue; $L=1$; $\alpha = 0.2$) for QFT on qubit-registers of varying sizes ($n$ as indicated). }
	\label{qsize}
\end{figure} 

To analyze the performance of PPOQC in larger systems, we implement Quantum Fourier Transform (QFT), which is central to several important quantum algorithms \cite{nielsen2007optimal}.  We implement the entire $n$-qubit QFT circuit, consisting of $n$ local and $O(n^2)$ conditional gates, into a single PP-Krotov GC sequence. 
The results, with registers up to seven qubits, shown in Fig. \ref{qsize} assure that PPOQC advantage persists even in larger systems. Further discussions and numerical analysis are provided in  supplemental materials.  In the following, we switch to an experimental implementation of PPOQC pulse-sequence.

\begin{figure}[t]
	\centering
	\includegraphics[trim=2.8cm 4.8cm 3.3cm 2.8cm,width=6.5cm,clip=]{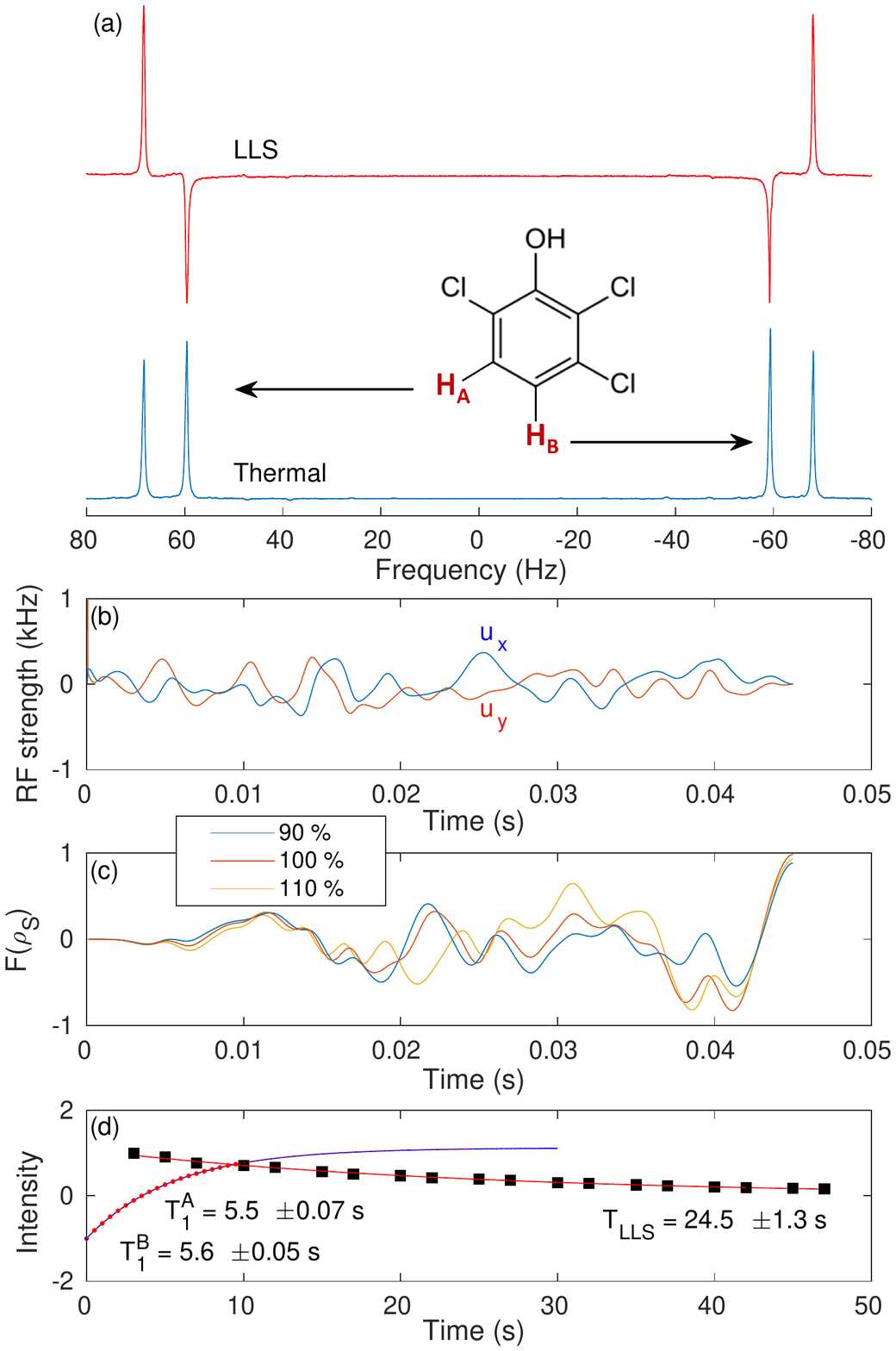}
	\caption{(a) Thermal and LLS spectra of TCP (molecule in inset). (b) PP-Krotov SC sequence ($L = 5$) preparing LLS directly from the thermal state.  (c) LLS fidelity evolution during the sequence in (b) at different RF inhomogeneity levels.  (d) $T_1$ values measured by the inversion recovery experiment and the $T_{LLS}$ measured by storage under spin-lock.}
	\label{tcpres}
\end{figure} 

\textit{NMR experiments:}
We now study the efficacy of PPOQC via an important application in NMR spectroscopy, i.e., preparation of a long-lived state (LLS).  Carravetta \textit{et al}. had demonstrated that the singlet-order of a homonuclear spin-pair outlives the usual life-times imposed by spin-lattice relaxation time constant ($T_1$)  \cite{carravetta2004beyond, carravetta2004long}.  Prompted by numerous applications in spectroscopy and imaging, several efficient ways of preparing LLS have been explored \cite{pileio2017singlet}. In the following, we utilize PP-Krotov SC optimization for this purpose.

We prepare LLS on two protons of 2,3,6-trichlorophenol (TCP; see Fig. \ref{tcpres} (a)).  Sample consists of 7 mg of TCP dissolved in 0.6 ml of deuterated dimethyl sulfoxide.  The experiments are carried out on a  Bruker 500 MHz NMR spectrometer at an ambient temperature of 300 K. Standard NMR spectrum of TCP shown in Fig. \ref{tcpres} (a) indicates resonance offset frequencies $\pm\Delta\nu/2$ to be $\pm 63.8$ Hz and the scalar coupling constant ${\tt J} = 8.8$ Hz.  The internal Hamiltonian of the system, in a frame rotating about the direction of the Zeeman field at an average Larmor frequency is 
$$H_0 = -\pi\Delta\nu I_z^A + \pi\Delta\nu I_z^B + 2\pi {\tt J} I_z^AI_z^B,$$
where $I_z^A$ and $I_z^B$ are the $z$-components of the spin angular momentum operators $\bf{I}^A$ and $\bf{I}^B$ respectively.  

The thermal equilibrium state at high-field and high-temperature approximation is of the form $\rho_0 = I_z^A + I_z^B$ (up to an identity term representing the background population).  
The goal is to design an RF sequence $\{u_x(t),u_y(t)\}$ introducing a time-dependent Hamiltonian
$$H(t) = H_0 + u_x(t) (I_x^A+I_x^B) + u_y(t) (I_y^A+I_y^B)$$
that efficiently transfers $\rho_0$ into zero-quantum singlet-triplet order $-\bf{I}^A\cdot \bf{I}^B$.  Under an RF spin-lock the triplet order decays rapidly while the singlet order $\rho_\mathrm{LLS}$ remains long-lived.  
The PP-Krotov SC pulse-sequence shown in Fig. \ref{tcpres} (b) consists of 1000 segments in a total duration of 45 ms, which is 30\% shorter than the standard sequence that requires $\frac{1}{2J}+\frac{3}{4\Delta\nu} = 63$ ms \cite{carravetta2004long}.  The fidelity profile shown in Fig. \ref{tcpres} (c) indicates the robustness of the sequence against $10\%$ RF inhomogeneity distribution with an average final fidelity above 95\%.  The LLS spectrum shown in Fig. \ref{tcpres}(a) is the characteristic of the singlet state $\rho_S$.
Fig. \ref{tcpres} (d) shows the experimental results of LLS storage under 1 kHz WALTZ-16 spin-lock.  It confirms the long life-time $T_{LLS}$ of about 24.5 s or about 4.5 times $T_1^A$ and $T_1^B$ measured by inversion recovery experiments.  A comparison with the standard method (as in ref. \cite{carravetta2004long}) revealed 27\% higher singlet order, further indicating the superiority of the PP-Krotov SC sequence.

\textit{Summary:}
At the heart of optimization algorithms lies a performance function that evaluates a process in relation to a  target.
Using a hybrid objective function that simultaneously takes into account a given target operator as well as a set of orthogonal operators we devised the \textit{push-pull} optimization of quantum controls. Combined influences of these operators not only results in a  faster convergence of the optimization algorithm, but also effects a better exploration of the parameter space and thereby generates better solutions.
Although the orthogonal set grows  exponentially with the system size, it is not necessary to include an exhaustive  set.  Even a small set of orthogonal operators, generated randomly during the iterations, can bring about a significant improvement in convergence.
While the push-pull approach can be implemented in a wide variety of quantum control routines, we described adopting it into a gradient based as well as a variational-principle based optimizations.  We observed considerable improvements in the convergence rates, without overburdening computational costs.
The numerical analysis with up to seven qubits confirmed that push-pull method retained superiority even in larger systems.
Finally, using NMR methods, we experimentally verified the robustness of a push-pull Krotov control sequence preparing a long-lived singlet order.  
Further work in this direction includes adaptive push-weights, optimizing the  functional forms of orthogonal gradients, generalization to open quantum controls, and so on. 
	 
\textit{Acknowledgments:}
Discussions with  Sudheer Kumar, Deepak Khurana, Soham Pal, Gaurav Bhole, Dr. Hemant Katiyar, and Dr. Pranay Goel are gratefully acknowledged.  This work was partly supported by DST/SJF/PSA-03/2012-13 and CSIR 03(1345)/16/EMR-II.

\bibliography{ref_newcontrol.bib}


\onecolumngrid
\section*{Supplemental information}
\begin{figure}[h]
	\centering
	\includegraphics[trim=8cm 8cm 8cm 3cm,width=6cm,clip=]{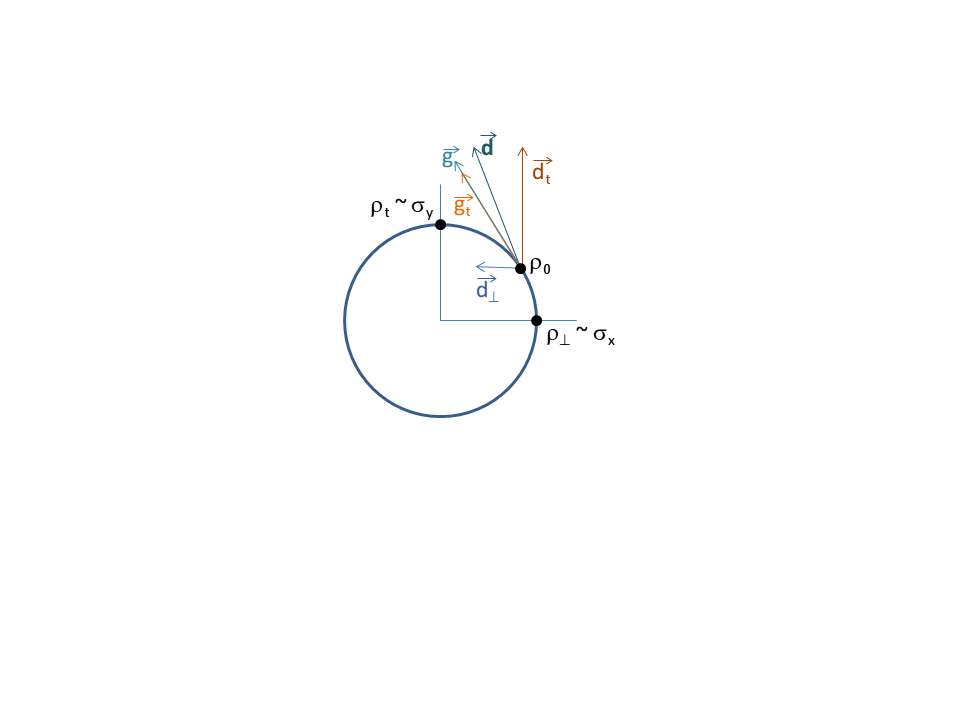}
	\caption{A naive model illustrating push-pull gradient being stronger than pull-only gradient.}
	\label{model}
\end{figure}

\textit{A naive model:}
Consider a single qubit state control problem with target being the pure state 
$\rho_t = (\mathbbm{1}+\sigma_y)/2$ and its orthogonal state being $\rho_\perp = (\mathbbm{1}+ \sigma_x)/2$. 
Consider an instantaneous state $\rho_0 = (\mathbbm{1}+\hat{n}\cdot \sigma)/2$. To simplify the picture, we consider the dynamics in xy-plane of the Bloch sphere by fixing $n_z = 0$ (see Fig. \ref{model}). In the pull-only scenario, the pull direction is along $\vec{d_t}$ that is parallel to $y$-axis.  Since the dynamics is constrained on the unit circle, the corresponding gradient $\vec{g_t}$ is the tangential component of $\vec{d_t}$.  In the push-pull case, we also have a direction $\vec{d_\perp}$ that is along $-x$-axis so that the net push-pull direction is along $\vec{d} = \vec{d_t}+\vec{d_\perp}$.  Now the corresponding tangential component $\vec{g}$ has a magnitude greater than $\vec{g_t}$ since $\vec{d}$ is the resultant of nonparallel vectors.  Of course, this simple model does not capture the entire picture, neither does it fully grasp the push-roles of orthogonal operators.  Nevertheless, the stronger gradients in the push-pull scenario hint about its faster convergence.

\begin{figure}[h]
	\centering
	\includegraphics[trim=4.8cm 10.9cm 5cm 11.3cm,width=9cm,clip=]{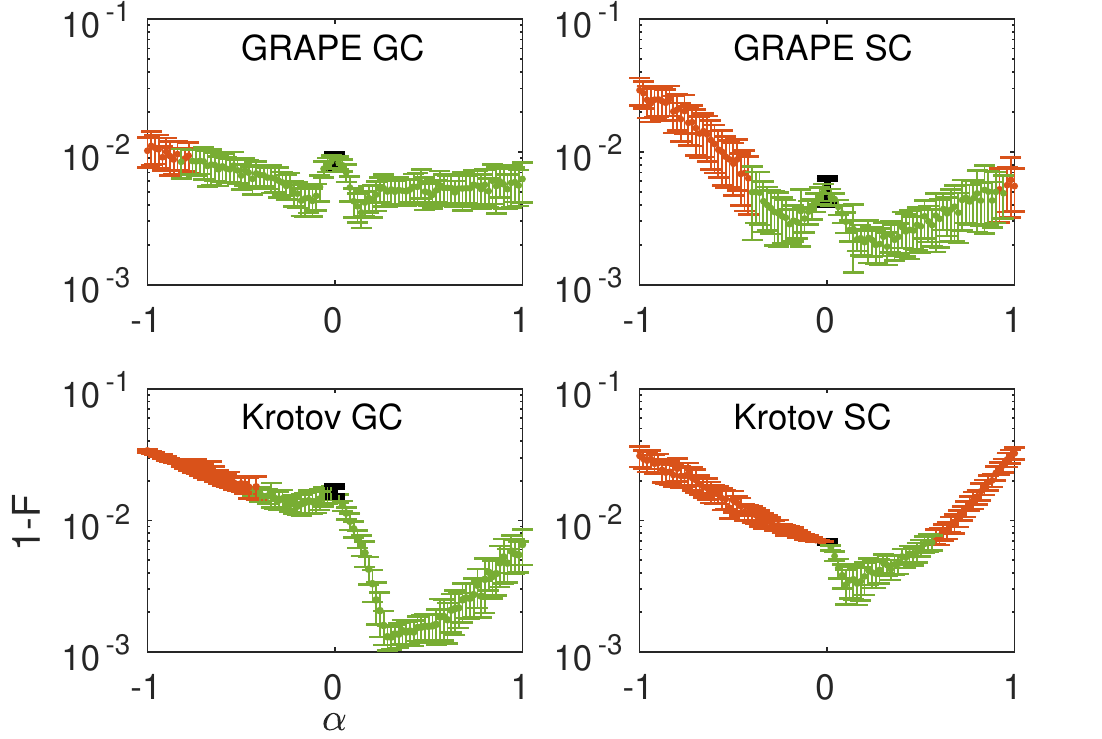}
	\caption{Infidelity versus the push-weight $\alpha$ for $L=6$.  Error bars indicate one standard deviation. The black point at $\alpha =0$ corresponds to the standard pull-only algorithms.  The green and red regions respectively indicate superior and inferior performances of PPOQC w.r.t. pull-only algorithm.}
	\label{alpha}
\end{figure}

\textit{Push-weight:}	
	Fig. \ref{alpha} displays infidelities of PP-GRAPE as well as PP-Krotov algorithms versus the push-weight $\alpha$.  We notice that, on the positive side, the infidelity is generally superior to the pull-only algorithm ($\alpha = 0$).  In each case, there exists an optimal push-weight roughly in the range $\alpha \in [0.1,0.3]$ at which the PPOQC works best. It is interesting to see that some negative regions also display superior performances.

\begin{figure}
	\centering
	\includegraphics[trim=0.3cm 0cm 0.5cm 0cm,width=9cm,clip=]{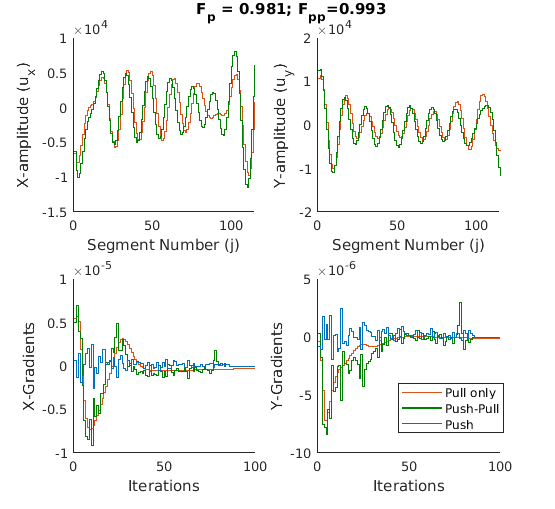}
	\caption{Top row: X and Y amplitudes for a two-qubit CNOT gate with pull-only GRAPE (red) and push-pull PP-GRAPE (green; $L=5$) algorithms.  Bottom row: Eovlution of X and Y gradients versus iteration number for one particular segment (segment number 78). Notice how the mean push gradients (blue) from the orthogonal operators modulate the effective push-pull gradients (green).}
	\label{parspace}
\end{figure}

\textit{Rapid parameter search in push-pull approach:}
To gain insight into the superiority of push-pull over pull-only approach, we observed how the gradients evolve over time.  Fig. \ref{parspace} displays the evolution of gradients versus control amplitudes over several  iterations.  The simulations are carried out for a two-qubit CNOT gate with both pull-only and push-pull GRAPE algorithms.  Push-pull algorithm ultimately converged to a better fidelity (0.993) than the pull-only algorithm (0.981).  Notice that the push-pull gradients show more rapid changes than the pull-only algorithm, indicating a more robust parameter search in action.  This behavior appears to be the crucial factor for the faster convergence of the push-pull approach.

\end{document}